# Deterministic single photon source enabled by coherent superposition of Mie-scattering moments in a NV- center coupled dipolar antenna

Faraz A. Inam[1]* and Rajesh V. Nair[2]†

*¹Department of Physics, Aligarh Muslim University, Aligarh, Uttar Pradesh, India*
*²Laboratory for Nano-scale Optics and Meta-materials (LaNOM)*
*Department of Physics, Indian Institute of Technology Ropar, Punjab 140001 India*

## Abstract

Generation of an ultra-bright, deterministic, solid-state single photon source with high photon collection rate is an imperative requirement for quantum technologies. In this direction, various nanophotonic systems coupled with single quantum emitters are being implemented, but results in low decay rate enhancement and MHz photon collection rate. Here, we unravel coherent superposition of excited Mie-scattering moments in a dipolar antenna, coupled with a single nitrogen-vacancy (NV-) center, to achieve bright single photon source with GHz collection rate. Such balancing of Mie-scattering moments, especially higher-order multi-polar moments, provide strong forward light scattering with null backward scattering at the generalized Kerker condition. This results in strong field intensity localization that can be used to shape the emission from an embedded NV- center in the dipolar antenna. A relative decay rate enhancement of more than 300 times with collection efficiency exceeding 75% is achieved that result in photon collection rate of ~ 5 GHz. The calculated intensity-intensity correlation confirms bright single photon emission with enhanced rate and collection efficiency.

**Keywords:** Mie-scattering, Kerker effect, quantum photonics, emission enhancement, NV centers, nanowires

* faraz.inam.phy@amu.ac.in

† rvnair@iitrpr.ac.in

## I. INTRODUCTION

The ultimate success of quantum technologies, in the second quantum revolution, like quantum key distribution and quantum information processing requires on-demand bright single-photon sources (SPS) with gigahertz (GHz) photon collection rate [1–3]. The SPS should have high purity, stable, room-temperature operable and indistinguishable. Many solid-state quantum emitters like defects in crystalline materials is emerging as SPS due to their photo-stability and high quantum efficiency [4–6]. The brightest known sources among these SPS have photon collection rates limited to few tens of megahertz, smaller than the required GHz rates [2]. Therefore, present day quantum technological applications use SPS based on spontaneous parametric down-conversion processes which provide indeterministic photon emission [2]. However, the requirement is on-demand SPS with 100% efficiency (*i.e.* whenever the source is triggered, a photon is generated) and high photon collection rate.

To increase the photon generation rate and subsequently, the photon collection rate from deterministic solid-state SPS, various resonator schemes such as photonic cavities [7], plasmonic nano-resonators [8], or hyperbolic meta-materials [9] are explored. The photon collection rate of 20 Mcps is achieved from single negatively-charged nitrogen-vacancy (NV⁻) center in nanodiamond integrated to a plasmonic patch antenna resonator [8]. Effectiveness of an antenna resonator is determined in terms of how efficiently it is able to enhance photon emission into the numerical aperture (NA) of collection lens and shape far-field radiation pattern of an emitter [10]. The collection efficiency (CE) which is defined as the fraction of total emitted power falling within the collection lens NA for plasmonic patch antenna resonator is observed to be less than 50% due to inherent plasmonic losses [8]. However, for applications in linear quantum computing, a theoretical study has set a minimum threshold CE value to be 2/3 (~ 67%) of total emission [11,12]. Therefore, an ultra-bright solid-state SPS with GHz photon emission rate with CE greater than 67% is required for sustained quantum technology.

When a resonator is placed in an electromagnetic field induces the excitation of electric dipole, magnetic dipole, electric quadrupole, magnetic quadrupole, and higher order moments, generally known as the Mie-scattering moments [13]. Mie-scattering moments can deterministically tune the local density of optical states (LDOS) which governs complete radiation process for an emitter [13–16]. When the first-order Mie-scattering moments such as electric and magnetic dipole are balanced, a complete forward light scattering is achieved, *aka*, the Kerker condition [17]. However, when all Mie-scattering moments (dipolar and quadrupole moments) are balanced, forward light scattering directionality is significantly enhanced, known as generalised Kerker condition [18,19]. The coherent superposition of Mie-scattering moments provides directional scattering [19,20], high-Q super-cavity modes [21], bound states in continuum [22], and LDOS enhancement using metasurfaces [23].

Here, we study deterministic coupling of NV- center to Mie-scattering moments of a silver (Ag) based coupled-dipolar antenna at the generalised Kerker condition. Strong directional light scattering is achieved, in the forward direction, along with field intensity confinement in the antenna. The NV- center is sandwiched between two identical Ag cylinders to form the coupled-dipolar antenna and sub-wavelength spacing forms plasmonic gap modes. The effective mixing of NV- center emission with Mie-scattering moments provide enhanced emission directionality with 80% CE value at 680 nm. A relative decay rate enhancement of 300 times together with CE of 80% provide a photon collection rate $\geq$ 5 GHz.

## II. RESULTS AND ANALYSIS

### IIA. COMPLETE FORWARD LIGHT SCATTERING

When the resonator size is of the order of the incident wavelength, far away from long-wavelength limit, higher-order multi-polar Mie-scattering moments with distinct radiation patterns are excited [24]. The far-field scattering pattern would be the superposition of individually excited Mie-scattering moments. The far-field differential scattering cross-section of the multi-polar Mie-scattering moments is given as [25]:

$$\frac{d\sigma^{scatt}}{d\Omega} = \frac{1}{k_0^2} \left[ \cos^2 \emptyset \left| S_{||}(\theta) \right|^2 + \sin^2 \emptyset |S_\perp(\theta)|^2 \right], \quad (2)$$

with $S_\perp(\theta)$ and $S_{||}(\theta)$ are the polarized scattering waves perpendicular and parallel to scattering plane:

$$S_{||}(\theta) = \sum_{n=1}^{n=\infty} \frac{2n+1}{n(n+1)} \left[ a_n \frac{dP_n^{(1)}(\cos\theta)}{d\theta} + b_n \frac{P_n^{(1)}(\cos\theta)}{\sin\theta} \right], \quad (3a)$$

$$S_\perp(\theta) = \sum_{n=1}^{n=\infty} \frac{2n+1}{n(n+1)} \left[ b_n \frac{dP_n^{(1)}(\cos\theta)}{d\theta} + a_n \frac{P_n^{(1)}(\cos\theta)}{\sin\theta} \right]. \quad (3b)$$

Here $\theta$ and $\phi$ are the polar and azimuthal angles, $P_n^{(1)}(\cos\theta)$ is the associated Legendre polynomial of degree 1, $a_n$ and $b_n$ are electric and magnetic Mie-scattering moments. The incident light propagates in +z direction with electric field oriented in x-direction. In general, the dominant excitations are lowest four multipolar moments, *i.e.* $a_1$ (electric dipole, ED), $b_1$ (magnetic dipole, MD), $a_2$ (electric quadrupole, EQ), and $b_2$ (magnetic quadrupole, MQ). Fig. 1(a) and 1(b) shows the coherent superposition of various Mie-scattering moments obtained through analytical calculations using Eq. 2. When only first-order moments are present ($a_1 = b_1$), we obtain $S_\parallel(\pi) = S_\perp(\pi) = 0$ that result in first Kerker condition (broken curve in Fig. 1(a)) [18]. When all individual moments are identical ($a_1 = b_1 = a_2 = b_2$), the superimposed scattering patterns shows complete forward scattering with strong directionality (solid curve) as shown in Fig. 1(a), known as generalised Kerker condition [18,19]. The balancing of the multipolar moments leads to complete phase symmetry with in-plane and out-of-plane scattering patterns being identical. When the scattering moments are of the same order but not identical ($a_1 = b_1 = a_2 = 1$ and $b_2 = 2$), which implies near balancing of multipolar moments (Fig.1(b)). The corresponding in-plane (solid curve) and out-of-plane (broken curve) scattering patterns are nearly same with substantial phase symmetry and strong forward directionality as the condition is close to generalised Kerker condition.

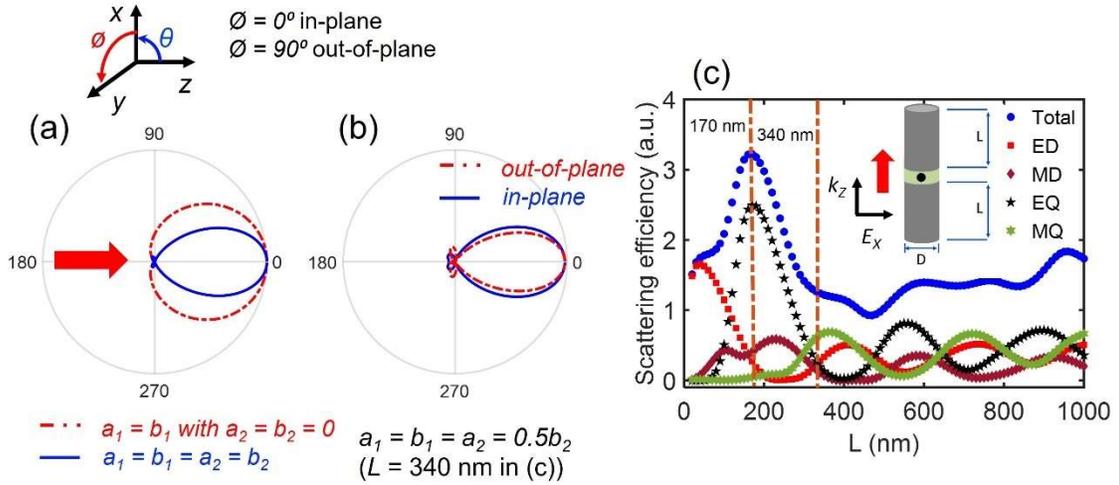

FIG. 1. Calculated far-field scattering pattern for multipolar Mie-scattering moments: (a) first Kerker condition (broken curve) and generalised Kerker condition (solid curve). (b) situation close to generalised Kerker condition with near balancing of multipolar moments. The incident plane wave propagates from left to right (z-direction, red arrow) with in-plane electric field along $x$-direction. (c) The individual Mie-scattering moments of the coupled dipolar antenna as a function of Ag cylinder's length ($L$) for longitudinal plane-wave excitation ($\hat{k}_z$). Vertical lines correspond to $L = 170$ and 340 nm, respectively. (Inset): Schematic of Ag based coupled-dipole antenna with nanodiamond based emitter sandwiched in the middle PVA layer.

The total scattering efficiency in-terms of individual multipolar Mie-scattering moments of a scatterer is written as [24]: $C_s^t = C_s^p + C_s^m + C_s^{Q^e} + C_s^{Q^m} + \cdots$

$$C_s^t = \frac{k^4}{6\pi\epsilon_0|E|^2}\left[\sum_\alpha\left(|p_\alpha|^2 + \frac{|m_\alpha|^2}{c}\right) + \frac{1}{120}\sum_{\alpha\beta}\left(|kQ_{\alpha\beta}^e|^2 + \frac{|kQ_{\alpha\beta}^m|^2}{c}\right) + \cdots\right]. \quad (4)$$

Here, $p_\alpha$ and $m_\alpha$ are ED and MD moments and $Q^e_{\alpha\beta}$ and $Q^m_{\alpha\beta}$ are EQ and MQ moments, respectively. $|E|$ is incident plane-wave amplitude, $k$ being wave-vector and $c$ vacuum light speed. The schematic of coupled dipolar antenna embedded with nanodiamond (black dot) containing single NV- center in a PVA layer is sandwiched between two identical Ag pillars of length ($L$) and diameter ($D$), as shown in Fig. 1(c) inset. The PVA layer thickness is 40 nm and nanodiamond size is 30 nm. The NV- center emission is centred at 680 nm and hence, physical parameters of coupled dipolar antenna are optimised for 680 nm in calculations. The nanodiamond NV- center is having quantum efficiencies of 0.1-0.9 [26] with an average efficiency being 0.7 [27]. Since, the PVA layer with nanodiamond is sandwiched between two identical Ag cylinders, a plasmonic nano-gap cavity is formed. This result in wavelength-dependent plasmonic gap modes for a plane-wave incidence through antenna with $D$ = 310 nm and $L$ = 340 nm. A broad plasmonic mode is obtained at $\lambda_0$ = 680 nm with a quality factor ($Q$) $\sim$ 30 (*Supplementary Fig. S1(a)*).

The scattering efficiency is calculated using electric field values at each mesh point in the computational grid under plane-wave excitation using Comsol Multi-physics module. Using these field values and permittivity profile at each mesh points, current density is calculated as: $J_\omega(r) = i\omega\epsilon_0(\epsilon_r - 1)E_\omega(r)$ [24]. Here $\epsilon_0$ and $\epsilon_r$ are permittivity of free space and Ag coupled-dipole antenna, respectively. The computationally obtained values of $E(r)$, $J_\omega(r)$, and $\varepsilon(r)$ are used to calculate individual multipolar Mie-scattering moments, $p_\alpha$, $m_\alpha$, $Q^e_{\alpha\beta}$ and $Q^m_{\alpha\beta}$ (*Supplementary material, SI*). Fig. 1(c) shows the calculated individual Mie-scattering moments for longitudinal ($\hat{k}_z$) excitation. The ED moment (squares) has a dominant contribution to total scattering efficiency for small $L$ ($<$ 100 nm). The higher-order multipolar moments become prominent for large $L$ value. For $L$=100-300 nm, the EQ moment (stars) has most dominant contribution with maxima obtained for $L$ = 170 nm. The total length of coupled dipolar antenna is 340 nm, for $L$ = 170 nm, which is equal to $\lambda/2$. The calculated wavelength-dependent scattering efficiency corroborates dominant contribution of EQ moment to scattering efficiency for $L$ = 170 nm (*Supplementary Fig. S1(b)*). The contribution from all four major Mie-scattering moments is of same order for $L\sim$ 340 nm, with ED = MD = EQ $\sim$ 0.3 and MQ $\sim$ 0.6. This corresponds to the regime of generalised Kerker condition ($a_1 = b_1 = a_2 = 0.5b_2$), similar to the case shown in Fig. 1(b). Therefore, far-field scattering pattern of antenna is expected to be highly directional. For transverse excitation ($\hat{k}_x$ or $\hat{k}_y$), the interaction between excitation field and antenna is less prominent (*Supplementary Fig. S1(c)*).

When we excite the quantum emitter (NV- center) embedded at the middle of coupled antenna, the emitted radiation is expected to generate multipolar Mie-scattering moments. The far-field radiation pattern of quantum emitter originates from the mixing of point dipole emission and scattering patterns of excited Mie-scattering moments. Fig. 2 shows the mixing of radiation pattern from Mie-scattering moments on the point dipole emission corresponds to $L = 170$ nm and 340 nm for longitudinal excitation. Since, the quantum emitter is emitting at the centre of the antenna, both top-down and bottom-up (red arrows) emission would be present as shown in Fig. 2. The point-dipole is considered to have its dipole moment oriented in $z$-direction in-order to maximise the coupling between dipole emission and Mie-scattering moments.

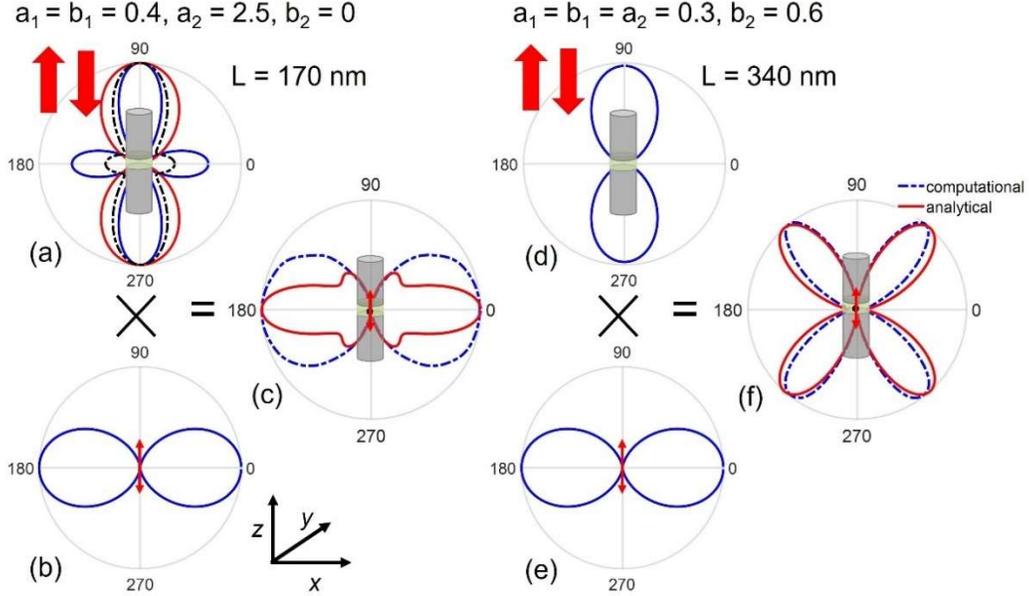

FIG. 2. The influence of the mixing of the far-field scattering pattern of the Mie-scattering moments on the final radiation pattern of the point dipole emission corresponding to two Ag cylinder's lengths, (a), (b) and (c) $L = 170$ nm and (d), (e) and (f) $L = 340$ nm, respectively.

We have seen in Fig. 1(c) that Mie-scattering moments are not balanced for $L = 170$ nm and hence, in-plane and out-of-plane far-field scattering patterns are expected to be different. The analytically calculated in-plane (blue curve) and out-of-pane (red curve) far-field scattering radiation patterns using Eq. 2 is given in Fig. 2(a). As the excitation source, the point-dipole has its dipole moment in $z$-direction and hence, radiation pattern due to Mie-scattering moments should be azimuthally symmetric. Further, both top-down and bottom-up excitations (red arrows in Fig. 2) should be present. The total scattering radiation pattern (dash-dotted curve) is therefore considered to be the average of above two cases (red and blue curves) as seen in Fig. 2(a). However, when all Mie-scattering moments are nearly balanced ($a_1 = b_1 = a_2 = 0.5$, $b_2 = 0.3$, Fig. 1(c)) for $L = 340$ nm, the in-plane and out-of-plane scattering patterns are nearly identical as seen in Fig. 1(b). The corresponding far-field scattering pattern for $L = 340$ nm is shown in Fig. 2(d). The radiation pattern for a quantum emitter in vacuum is shown in Fig. 2(b) and 2(e). The emission pattern of same quantum emitter embedded in the dipolar antenna with $L = 170$ nm and $L = 340$ nm is shown in Fig. 2(c) and 2(f), respectively. The agreement between calculated (red curve) and simulated (blue curve) emission pattern due to mixing of dipole emission with Mie-scattering moments is commendable for $L = 340$ nm. The modified emission pattern of quantum emitter shows more scattered intensity in the up-ward and down-ward direction. This implies that a proper reflecting surface beneath the antenna can significantly boost CE values from top direction. The CE value is obtained from top direction using a collection objective with NA = 0.9 and solid angle of $\sim 64°$ (*Supplementary material, SII*). The maximum CE value of 40% is achieved for $L \sim 340$ nm, when all Mie-scattering moments are nearly balanced ($a_1 = b_1 = a_2 = 0.3$ and $b_2 = 0.6$) as seen in Fig. 1(b). The CE value for $L = 170$ nm is expected to be much lower as the emission is mostly directed towards the sideways as seen in Fig. 2(c). For $L = 170$ nm, the Mie-scattering moments excited through transverse excitation are significant and expected to have an influence on emitter final radiation pattern (*Supplementary Fig. S1(c)*).

## IIB. EMISSION RATE MODIFICATION

We now study the influence of plasmonic gap-cavity induced modification on the emission decay rate of single NV- center. We calculate the relative decay rate which is defined as the total power radiated over a closed surface by the NV- center coupled antenna divided by the power radiated by NV- center in a reference sample. The nanodiamond with single NV$^-$ emitting in vacuum is taken as reference sample. The coupled dipolar antenna is excited by a plane-wave and maximum field localization is obtained at cavity centre. Such localized field intensity lead to large LDOS enhancement for NV- center placed in the cavity following Eq. S2 (*Supplementary material, SII*), which in-turn would lead to decay rate enhancement. Fig. 3(a) shows the calculated relative decay rate (circles) of single NV- center coupled to antenna of $L = 340$ nm emitting at 680 nm as a function of $D$. The first and second resonant mode appear at $D = 310$ nm and 620 nm, respectively. The relative decay rate enhancement is corroborated with enhanced field values (squares) in the cavity layer as seen in Fig. 3(a).

In-order to maximise the relative decay rate as well as CE from top direction, the single emitter coupled dipolar antenna is placed on top of a 100 nm thick Ag layer. Fig. 3(b) shows calculated relative decay rate (circles) and CE values (squares) as a function of $L$. The periodic behaviour in the decay rate curve is originated due to in-phase reflections occurring from bottom reflector [14]. The total decay rate enhancement is observed to remain well over two-orders of magnitude throughout the studied $L$ values. It is seen that the maximum decay rate enhancement is accompanied with minimum CE value of 10% for $L = 170$ nm, which support emission leakage to sideways as seen in Fig. 2(c). The CE value reaches a maximum value of 80% for $L$ = 340 nm which is induced by the balancing of Mie-scattering moments. The emission pattern with Ag reflector layer beneath the antenna with $L$ = 340 nm is shown in Fig. 4(a). The bottom Ag reflector layer is directing most of the emission towards the top direction. The decay-rate enhancement is observed to be around 400 times with CE value of 80%. On averaging over entire NV$^-$ center emission (630-750 nm), the enhancement is found to be 300 with a CE of 75%. The wavelength-dependent relative decay rate enhancement as well as CE values shows the maximum at 680 nm for nanoantenna with $L$ = 340 nm (*Supplementary material, SII*).

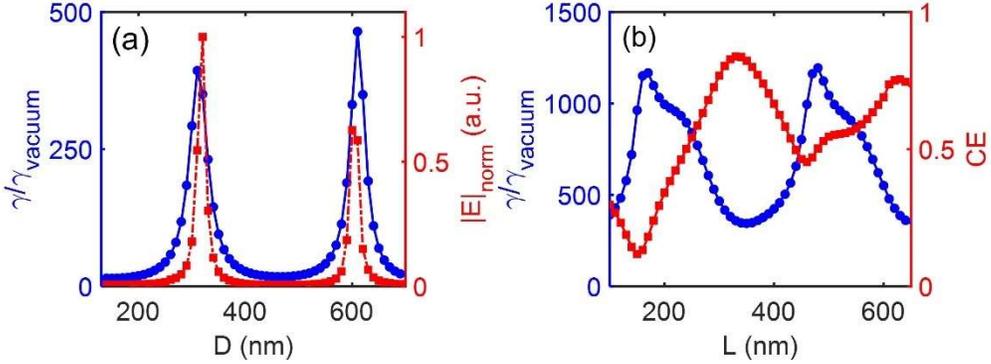

FIG. 3. (a) The normalized electric field (squares) at the cavity centre along with relative decay rate (circles) at 680 nm as a function of Ag cylinder's diameter $D$ with $L$ = 340 nm. The dipole moment is along the z-direction (b) The variation of relative decay rate (circles) together with CE (squares) as a function of $L$ for the NV- center coupled antenna placed on a reflecting layer.

Using the decay-rate enhancement and CE values, the photon collection rate can be estimated. Suppose NV$^-$ center is placed in a medium with an index 1.05 close to that of vacuum. The average lifetime ($\tau_0$) in such a medium is 31 ns with an average intrinsic quantum efficiency (QE$_i$) of 0.7 [27]. Such a SPS is expected to emit photons at a rate $\sim QE_i \times \frac{1}{\tau_0} = 0.7 \times \frac{1}{31 \times 10^{-9} \, ns} \sim 22$ MHz. In the proposed NV- center coupled antenna, average decay rate enhancement obtained over NV$^-$ center spectrum is 300. Hence, photon emission rate would be enhanced to $300 \times 22 \times 10^6 = 6.6$ GHz. With an average CE $\sim 0.75$ over the NV$^-$ center spectrum, photon collection rate should be: $0.75 \times 6.6 \sim 5$ GHz, which is well into the required GHz regime for quantum technology. A comparison between the results achieved in the present work with other single emitter coupled nanophotonic schemes suggests a better collection efficiency for the proposed coupled dipolar antenna (Table 1 in Methods).

## IIC. SINGLE PHOTON EMISSION STUDIES

We have further calculated the second-order intensity-intensity correlation function ($g^2(\tau)$) which defines statistics of light emission with $\tau$ represents time-delay between intensity correlation events [28]. The $g^{(2)}(\tau)$ is defined as: $g^{(2)}(\tau) = \frac{\langle n_1(t)n_2(t+\tau)\rangle}{\langle n_1(t)\rangle\langle n_2(t+\tau)\rangle}$, with $n_i(t)$ is number of counts registered on detector $i$ at time $t$ and $\tau$ defines the time separation between the photon detection events on the two detectors [49]. A realistic SPS is expected to emit photons at separate time intervals defined by the excitation source rate, with no correlations between different photon bursts so that $g^{(2)}(\tau = 0) = 1 - \frac{1}{n}$. We use a two-level system to calculate $g^{(2)}(\tau)$ for a single emitter coupled to the proposed antenna [29]. For a two-level system driven by off-resonant excitation, $g^{(2)}(\tau) = 1 - e^{-\frac{\tau}{\tau_1}}$ with $\tau_1$ being anti-bunching decay time. Here, $\tau = 0$ implies $g^{(2)}(\tau)$ = 0, such a system behaves as a SPS. For very low excitation power ($P \to 0$), the pump rate coefficient $k_{12} = 0$ and $\tau_1 \approx \frac{1}{k_{21}}$, $k_{21}$ being the excited state's decay rate. As $\frac{1}{k_{21}} = \tau_0$, $\tau_0$ is the emission lifetime of the two-level emitter [30], for very low excitation power ($P \to 0$), $\tau_1 \approx \tau_0$, anti-bunching decay time $\tau_1$ approaches emitter lifetime $\tau_0$. An emission decay rate enhancement would lead to reduction in emitter lifetime $\tau_0$ with subsequent reduction in anti-bunching decay time $\tau_1$ ($P \to 0$).

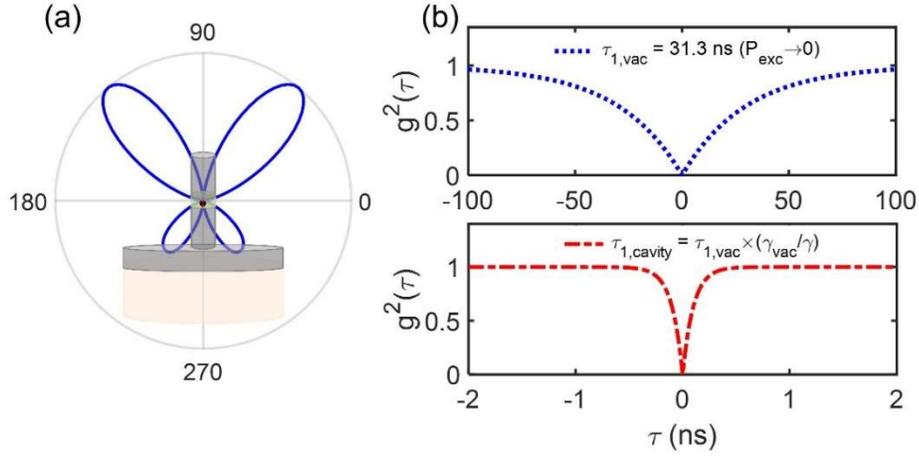

FIG. 4. (a) The far-field radiation pattern of a NV- center coupled antenna with $L$ = 340 nm placed on a reflector layer. (b) The second-order intensity-intensity correlation function ($g^{(2)}(\tau)$) for the single NV$^-$ center emission in vacuum (top: dotted curve) and in coupled dipolar antenna (bottom: dash-dotted curve).

Figure 4(b) shows calculated $g^{(2)}(\tau)$ for NV- center emitting in vacuum (dotted curve) and in the proposed coupled dipolar antenna (dash-dotted curve), respectively. The $g^{(2)}(\tau)$ curve width centred around $\tau = 0$ is drastically reduced by 300 times for NV- center coupled to the antenna. This shows 300 times average decay rate enhancement. The anti-bunching decay time, $\tau_{1,cav}$ ($P \to 0$) = $\tau_{1,vac} \times (\gamma_{vac}/\gamma)$ = $\tau_{1,vac}/300$. Thus, we can generate bright single photons from single NV- center coupled with Ag antenna with GHz photon collection rate. Our results put forward an amenable, easy-to-fabricate antenna for generating single photons with high emission as well as collection rate, which can induce paradigm shift in NV- center based quantum technologies. The proposed scheme is not limited to NV- center based SPS but the concept is applicable to other solid-state single emitters.

## III. CONCLUSIONS

The electromagnetic multi-polar Mie-scattering moments associated with an antenna can tune the far-field radiation pattern of the coupled quantum emitters. The coherent superposition of excited Mie-scattering moments in a coupled dipolar antenna results in generalized Kerker condition with complete forward light scattering. Such coherent superposition of multi-polar moments modifies the emission pattern of the embedded single NV- center at its maximum emission wavelength of 680 nm. A relative decay rate enhancement of more than 300 times with a collection efficiency of ~75% is achieved using the proposed antenna. The rate enhancement achieved in the present work result in 5 GHz photon collection rates, which is a sought-after goal for implementing quantum technologies. Our results are therefore substantial to accelerate research for generating on-demand single photon source with GHz collection rate for quantum technological applications.

### METHODS

**Numerical Calculations.** The finite-element method (FEM) simulations are done using commercial software package (Comsol Multiphysics RF Module). The Mie scattering moments and radiation patterns are simulated using FEM method. Analytical calculation of intensity-intensity correlation is done using a custom-written Matlab code.

Table I. Comparative study of emission enhancement reported from a quantum emitter based room-temperature stable, solid-state SPS. (cps = counts per second)

| Photonic Structures | Diamond nanowire [7] | Diamond parabolic mirror [8] | "Bullseye" diamond grating [9] | Ag nano-patch antenna [10] | Ag nanopatch antenna [11] | present manuscript |
|---|---|---|---|---|---|---|
| Active system | NV− center | NV− center | NV− center | NV− center | CdSe/ZnS QD | NV− center |
| Decay rate enhancement | ~1 | ~1 | ~1 | 70 | 540 | 300 |
| Collection efficiency | 0.4 | 0.75 | 0.35 | 0.5 | 0.42 | 0.75 |
| Photon collection rate | 200 Kcps | 4 Mcps | 3 Mcps | 20 Mcps | 1 Mcps | few Gcps |

**Notes:**

The authors declare no competing financial interest

**ACKNOWLEDGMENTS**

The authors would like to acknowledge the financial support from the Department of Science and Technology (DST), India (CRG/2021/001167). RVN acknowledges the financial support from DST-ICPS [DST/ICPS/QuST/Theme-2/2019/General], DST-SERB [SB/SJF/2020-21/05], and the Swarnajayanti Fellowship (DST/SJF/PSA-01/2019-20). The authors also thank Dr. Nadeem Ahmed and Dr. Megha Khokhar for useful discussions during the course of this study.

———————————————

[1]    S. Scheel, *Single-Photon Sources–an Introduction*, J. Mod. Opt. **56**, 141 (2009).

[2]    I. Aharonovich, D. Englund, and M. Toth, *Solid-State Single-Photon Emitters*, Nat. Photonics **10**, 631 (2016).

[3]    M. Atatüre, D. Englund, N. Vamivakas, S. Y. Lee, and J. Wrachtrup, *Material Platforms for Spin-Based Photonic Quantum Technologies*, Nat. Rev. Mater. **3**, 38 (2018).

[4]    L. Schlipf et al., *A Molecular Quantum Spin Network Controlled by a Single Qubit*, Sci. Adv. **3**, e1701116 (2017).

[5]    I. Aharonovich, S. Castelletto, D. A. Simpson, C.-H. Su, A. D. Greentree, and S. Prawer, *Diamond-Based Single-Photon Emitters*, Reports Prog. Phys. **74**, 076501 (2011).

[6]    T. T. Tran, K. Bray, M. J. Ford, M. Toth, and I. Aharonovich, *Quantum Emission from Hexagonal Boron Nitride Monolayers*, Nat. Nanotechnol. **11**, 37 (2015).

[7]    T. M. Babinec, B. J. M. Hausmann, M. Khan, Y. Zhang, J. R. Maze, P. R. Hemmer, and M. Lončar, *A Diamond Nanowire Single-Photon Source*, Nat. Nanotechnol. **5**, 195 (2010).

[8]    S. I. Bogdanov et al., *Ultrabright Room-Temperature Sub-Nanosecond Emission from Single Nitrogen-Vacancy Centers Coupled to Nanopatch Antennas*, Nano Lett. **18**, 4837 (2018).

[9]    A. Kala, F. A. Inam, S. Biehs, P. Vaity, and V. G. Achanta, *Hyperbolic Metamaterial with Quantum Dots for Enhanced Emission and Collection Efficiencies*, Adv. Opt. Mater. **8**, 2000368 (2020).

[10]   A. F. Koenderink, *Single-Photon Nanoantennas*, ACS Photonics **4**, 710 (2017).

[11]   M. Varnava, D. E. Browne, and T. Rudolph, *How Good Must Single Photon Sources and Detectors Be for Efficient Linear Optical Quantum Computation?*, Phys. Rev. Lett. **100**, 060502 (2008).

[12]   E. Knill, R. Laflamme, and G. J. Milburn, *A Scheme for Efficient Quantum Computation with Linear Optics*, Nature **409**, 46 (2001).

[13]   M. Kerker, *The Scattering of Light and Other Electromagnetic Radiation* (Academic Press, New York, 1969).

[14]   W. L. Barnes, S. A. R. Horsley, and W. L. Vos, *Classical Antennas, Quantum Emitters, and Densities of Optical States*, J. Opt. **22**, 073501 (2020).

[15]   L. Novotny and B. Hecht, *Principles of Nano-Optics* (Cambridge University Press, 2012).

[16]   C. F. Bohren and D. R. Huffman, *Absorption and Scattering of Light by Small Particles*

(Wiley, 1998).

[17] S. Person, M. Jain, Z. Lapin, J. J. Sáenz, G. Wicks, and L. Novotny, *Demonstration of Zero Optical Backscattering from Single Nanoparticles*, Nano Lett. **13**, 1806 (2013).

[18] W. Liu and Y. S. Kivshar, *Generalized Kerker Effects in Nanophotonics and Meta-Optics*, Opt. Express **26**, 13085 (2018).

[19] R. Alaee, R. Filter, D. Lehr, F. Lederer, and C. Rockstuhl, *A Generalized Kerker Condition for Highly Directive Nanoantennas*, Opt. Lett. **40**, 2645 (2015).

[20] K. Koshelev and Y. Kivshar, *Dielectric Resonant Metaphotonics*, ACS Photonics **8**, 102 (2020).

[21] M. V. Rybin, K. L. Koshelev, Z. F. Sadrieva, K. B. Samusev, A. A. Bogdanov, M. F. Limonov, and Y. S. Kivshar, *High- Q Supercavity Modes in Subwavelength Dielectric Resonators*, Phys. Rev. Lett. **119**, (2017).

[22] Z. Liu, Y. Xu, Y. Lin, J. Xiang, T. Feng, Q. Cao, J. Li, S. Lan, and J. Liu, *High- Q Quasibound States in the Continuum for Nonlinear Metasurfaces*, Phys. Rev. Lett. **123**, (2019).

[23] M. Khokhar, F. A. Inam, and R. V. Nair, *Kerker Condition for Enhancing Emission Rate and Directivity of Single Emitter Coupled to Dielectric Metasurfaces*, Adv. Opt. Mater. **10**, 2200978 (2022).

[24] R. Alaee, C. Rockstuhl, and I. Fernandez-Corbaton, *An Electromagnetic Multipole Expansion beyond the Long-Wavelength Approximation*, Opt. Commun. **407**, 17 (2018).

[25] A. E. Miroshnichenko, J. Y. Lee, and R.-K. Lee, *Simultaneously Nearly Zero Forward and Nearly Zero Backward Scattering Objects*, Opt. Express, Vol. 26, Issue 23, Pp. 30393-30399 **26**, 30393 (2018).

[26] A. Mohtashami and A. Femius Koenderink, *Suitability of Nanodiamond Nitrogen–Vacancy Centers for Spontaneous Emission Control Experiments*, New J. Phys. **15**, 043017 (2013).

[27] F. A. Inam et al., *Emission and Nonradiative Decay of Nanodiamond NV Centers in a Low Refractive Index Environment*, ACS Nano **7**, 3833 (2013).

[28] R. Loudon, *The Quantum Theory of Light.*, 1st ed. (Clarendon Press, Michigan, USA, 1973).

[29] C. Kurtsiefer, S. Mayer, P. Zarda, and H. Weinfurter, *Stable Solid-State Source of Single Photons*, Phys. Rev. Lett. **85**, 290 (2000).

[30] S. Castelletto, F. A. Inam, S. I. Sato, and A. Boretti, *Hexagonal Boron Nitride: A Review of the Emerging Material Platform for Single-Photon Sources and the Spin-Photon Interface*, Beilstein J. Nanotechnol. **11**, 740 (2020).